\documentclass[aps,preprint,preprintnumbers,amsmath,amssymb,nofootinbib]{revtex4}
\usepackage{graphicx}
\usepackage{epsfig}
\newcommand {\beg}{\begin{equation}}
\newcommand {\en}{\end{equation}}
\newcommand {\bega}{\begin{eqnarray}}
\newcommand {\ena}{\end{eqnarray}}
\begin{document}
\title{Induced Matter Theory of gravity from a Weitzenb\"ock 5D vacuum and pre-big bang collapse of the universe}
\author{$^{2}$
Jes\'us Mart\'{\i}n Romero\footnote{E-mail address:
jesusromero@conicet.gov.ar}, $^{1,2}$ Mauricio Bellini
\footnote{E-mail address: mbellini@mdp.edu.ar} }
\address{$^1$ Departamento de F\'isica, Facultad de Ciencias Exactas y
Naturales, Universidad Nacional de Mar del Plata, Funes 3350, C.P.
7600, Mar del Plata, Argentina.\\
$^2$ Instituto de Investigaciones F\'{\i}sicas de Mar del Plata (IFIMAR), \\
Consejo Nacional de Investigaciones Cient\'ificas y T\'ecnicas
(CONICET), Argentina.}

\begin{abstract}
We extend the Induced Matter Theory of gravity (IMT) to 5D curved
spacetimes by using the Weitzenb\"ock representation of
connections on a 5D curved spacetime. In this representation the
5D curvature tensor becomes null, so that we can make a static
foliation on the extra noncompact coordinate to induce in the
Weitzenb\"ock representation the Einstein equations. Once we have
done it, we can rewrite the effective 4D Einstein equations in the
Levi-Civita representation. This generalization of IMT opens a
huge window of possible applications for this theory. A pre-big
bang collapsing scenario is explored as an example.
\end{abstract}
\maketitle

\section{Introduction and Motivation}

Pre-big bang cosmology is a new and fascinating cosmological
paradigm\cite{gas} that emerges from the possibility that the our
universe could be born from a unstable compact object (could be a
black-hole), with a low state of entropy. Under this point of view
the universe is just one of many in a cyclical chain, with each
Big Bang starting up a new universe in place of the one before,
and the end of each universe will involve a return to low entropy.
This is because black holes suck in all the matter, energy, and
information they encounter, which works to remove entropy from our
universe. However, the investigation of the black-holes formation
is not closed. The problem with this issue consists of explaining
how the gravitational collapse does not end with a singularity
with infinity energy density. This topic has been studied earlier
by Liu and Wesson\cite{5} in the framework of IMT. In particular,
the Induced Matter Theory (IMT) is based on the assumption that
ordinary matter and physical fields that we can observe in our 4D
universe can be geometrically induced from a 5D Ricci-flat metric
with a space-like noncompact extra dimension on which we define a
physical vacuum\cite{IMT,imt1}. In order to study the
gravitational collapse resulting in the formation of compact
objects with finite size and energy densities from a 5D vacuum
state, we shall incorporate the Weitzenb\"ock connections to
extend the Induced Matter Theory (IMT) formalism\cite{IMT}, which
is a very active extra dimensional theory. The induced matter
theory is regarded a non-compact KK theory in 5D, since the fifth
extra dimension is assumed extended. This theory is mathematically
supported by the embedding Campbell-Magaard theorem \cite{cam1,
cam2}. The main idea in this theory is that matter in 4D can be
geometrically induced from a 5D Ricci-flat metric. Thus, the
theory considers a 5D apparent vacuum defined by
$^{(5)}R_{AB}=0$\footnote{In our conventions indices "a,b,c,..,h"
run from $1$ to $5$, "A,B,C,..,H" run from $1$ to $5$, Greek
indices $\alpha,\beta,\gamma$ run from $1$ to $4$ and indices
"i,j,k,..." run from $2$ to $4$.}, which are the field equations
of the theory.

In this work we study an extension of the  IMT to using the
Weitzenb\"ock\cite{weit} representation of connections on a 5D
curved spacetime. In this representation the 5D curvature tensor
(which could be nonzero in a Levi-Civita representation) becomes
null in a Weitzenb\"ock one, so that we can make static foliation
on the extra noncompact coordinate to induce in the Weitzenb\"ock
representation the effective 4D Einstein-Cartan equations. Once we
have done it, we can rewrite the effective 4D Einstein equations
in the Levi-Civita representation. There are multiple possible
applications for this formalism. One of these applications is the
treatment of a pre-big bang collapsing universe. This issue is
studied in the Sect. III.

\section{Weitzenb\"ock connections and the extended IMT}

In this section we shall introduce the Weitzenb\"ock connections
to develop the IMT in the framework of the Weitzenb\"ock
treatment.

\subsection{Weitzenb\"ock treatment in 5D}

In order to extend the IMT formalism we shall use the
Weitzenb\"ock connections, which are defined when we make a
transformation
\begin{equation}
{\bf e}_a = e^A_{\,\,a} {\bf E}_A,
\end{equation}
from a coordinated basis ${\bf E}_A$ to another ${\bf e}_{a}$.
Hence, all tensors can be written using the vielbein $e^A_{\,\,a}$
and its inverse $\bar{e}^a_{\,\,A}$, such that, if $e^A_{\,\,a}
\bar{e}^b_{\,\,A} = \delta^b_a$ and
\begin{equation}
g_{ab} = e^A_{\,\,a}e^B_{\,\,b} g_{AB}.
\end{equation}
When the covariant derivative of the vielbein becomes zero:
$\bar{\nabla}_b \,e^A_{\,\,a}=0$, the connection with respect to
which we make the covariant derivative, is a Weitzenb\"ock
connection, $\bar{\Gamma}^c_{\,\,\,ba}$:
\begin{equation}
\bar{\nabla}_b \,e^A_{\,\,a}= \partial_b \,e^A_{\,\,a}-
\bar{\Gamma}^c_{\,\,\,ba}\,e^A_{\,\,c}=0.
\end{equation}
We shall consider that the curvature tensor of the 5D space
represented by the metric tensor $g_{AB}$ is zero from the point
of view of the Levi-Civita and Weitzenb\"ock connections. However,
the transformed space represented by tensor metric $g_{ab}$ is
Weitzenb\"ock-flat: $\bar{R}^a_{\,\,\,bcd}=0$, but not
Riemann-flat: ${R}^a_{\,\,\,bcd}\neq 0$. The curvature in the
Riemann and Weitzenb\"ock representations, is given (on the space
$g_{ab}$), by
\begin{eqnarray}
\bar{R}^a_{\,\,\,bcd} &= &\bar{\Gamma}^a_{\,\,\,bd,c} -
\bar{\Gamma}^a_{\,\,\,bc,d} + \bar{\Gamma}^e_{\,\,\,bd}
\bar{\Gamma}^a_{\,\,\,ec} - \bar{\Gamma}^e_{\,\,\,bc}
\bar{\Gamma}^a_{\,\,\,ed}=0, \label{r1} \\
{R}^a_{\,\,\,bcd} &= & {\Gamma}^a_{\,\,\,bd,c} -
{\Gamma}^a_{\,\,\,bc,d} + {\Gamma}^e_{\,\,\,bd}
{\Gamma}^a_{\,\,\,ec} - {\Gamma}^e_{\,\,\,bc}
{\Gamma}^a_{\,\,\,ed} \neq 0. \label{r2}
\end{eqnarray}
Here, the Weitzenb\"ock and Levi-Civita connections on $g_{ab}$,
are, respectively, given by
\begin{eqnarray}
\bar{\Gamma}^a_{\,\,\,bc} & = & \bar{e}^a_A \, \partial_c \,e^A_b, \\
{\Gamma}^a_{\,\,\,bc} & = & \frac{1}{2} g^{as} \left(\partial_{b}
g_{cs} + \partial_c g_{cs} - \partial_s g_{bc}\right),
\end{eqnarray}
such that both connections are related by
\begin{equation}\label{con}
\bar{\Gamma}^a_{\,\,\,bc} = {\Gamma}^a_{\,\,\,bc} -
\bar{K}^a_{\,\,\,bc},
\end{equation}
$\bar{K}^a_{\,\,\,bc}$ being the contortion in the Weitzenb\"ock
space, which is defined from the torsion
$\bar{T}^a_{\,\,\,bc}=\bar{\Gamma}^a_{\,\,\,bc}-\bar{\Gamma}^a_{\,\,\,cb}$
\begin{equation}
\bar{K}^a_{\,\,\,bc} = \frac{1}{2}
\left(\bar{T}^{\,\,a}_{b\,\,\,c}+\bar{T}^{\,\,a}_{c\,\,\,b}-\bar{T}^{a}_{\,\,\,bc}\right).
\end{equation}
By contracting the tensor $\bar{R}^a_{\,\,\,bcd}$, it is possible
to obtain two tensors, one symmetric
$\bar{R}_{bc}=\bar{R}^a_{\,\,\,bca}$, and the other antisymmetric,
$\bar{S}_{cd}=\bar{R}^a_{\,\,\,acd}$, which are
\begin{eqnarray}
\bar{R}_{bc} &= &\bar{\Gamma}^a_{\,\,\,ba,c} -
\bar{\Gamma}^a_{\,\,\,bc,a} + \bar{\Gamma}^e_{\,\,\,ba}
\bar{\Gamma}^a_{\,\,\,ec} - \bar{\Gamma}^e_{\,\,\,bc}
\bar{\Gamma}^a_{\,\,\,ea}=0, \label{rr1}\\
\bar{S}_{cd} &= &\bar{\Gamma}^a_{\,\,\,ad,c} -
\bar{\Gamma}^a_{\,\,\,ac,d}=0. \label{rr2}
\end{eqnarray}

\subsection{IMT with the Weitzenb\"ock representation}

Now we shall study the effective 4D Einstein equations after we
make a static foliation on the 5D Einstein equations using the
Weitzenb\"ock representation. We consider Eq. (\ref{rr1}). If we
take into account only the 4D components of the tensor
$\bar{R}_{ab}$, we obtain, after separately writting the sum over
the 4D and extra dimensional components
\begin{eqnarray}
\left.\overbrace{\bar{R}_{\beta\gamma}}^{5D}\right|_{l=l_0} &=
&\left.\overbrace{\bar{\Gamma}^{\alpha}_{\,\,\,\beta\alpha,\gamma}
- \bar{\Gamma}^{\alpha}_{\,\,\,\beta\gamma,\alpha} +
\bar{\Gamma}^\epsilon_{\,\,\,\beta\alpha}
\bar{\Gamma}^{\alpha}_{\,\,\,\epsilon\gamma} -
\bar{\Gamma}^{\epsilon}_{\,\,\,\beta\gamma} \bar{\Gamma}^{\alpha}_{\,\,\,\epsilon\alpha}}^{4D}\right|_{l=l_0} \nonumber \\
&+& \left[ \left(\bar{\Gamma}^5_{\,\,\,\beta 5,\gamma} -
\bar{\Gamma}^5_{\,\,\,\beta\gamma, 5}\right) \right. \nonumber \\
&+& \left.\left( \bar{\Gamma}^5_{\,\,\,5 \gamma}
\bar{\Gamma}^5_{\,\,\,\beta 5} + \bar{\Gamma}^5_{\,\,\,\mu\gamma}
\bar{\Gamma}^{\mu}_{\,\,\,\beta 5} +
\bar{\Gamma}^{\alpha}_{\,\,\,5 \gamma }
\bar{\Gamma}^5_{\,\,\,\beta \alpha} \right.\right. \nonumber \\
&-& \left.\left.\left.\bar{\Gamma}^5_{\,\,\,55}
\bar{\Gamma}^5_{\,\,\,\beta\gamma} - \bar{\Gamma}^5_{\,\,\,\mu 5}
\bar{\Gamma}^{\mu}_{\,\,\,\beta \gamma} -
\bar{\Gamma}^{\mu}_{\,\,\,5 \mu}
\bar{\Gamma}^5_{\,\,\,\beta\gamma}\right)
\right]\right|_{l=l_0}=0. \label{ind}
\end{eqnarray}
We shall restrict our study to consider canonical metrics: $dS^2=
\left(\frac{l}{l_0}\right)^2
h_{\alpha\beta}(y^{\gamma})\,dy^{\alpha} dy^{\beta}-dl^2$, where
$l$ is the noncompact extra dimension and $l_0$ is a constant
introduced by consistency of physical dimensions. It is possible
to demonstrate that
$\left.\overbrace{\bar{\Gamma}^\epsilon_{\,\,\,\beta\alpha}}^{5D}\right|_{l=l_0}
=\overbrace{\bar{\Gamma}^\epsilon_{\,\,\,\beta\alpha}}^{4D}$,
where $\overbrace{\bar{\Gamma}^\epsilon_{\,\,\,\beta\alpha}}^{4D}$
is the effective 4D connection of 4D hypersurface
$h_{\alpha\beta}$. The induced 4D Weitzenb\"ock tensor
$\overbrace{\bar{R}_{\beta\gamma}}^{4D}\equiv
\left.\overbrace{\bar{R}^{a}_{\beta\gamma a}}^{5D}\right|_{l=l_0}$
is given by
\begin{eqnarray}
\overbrace{\bar{R}_{\beta\gamma}}^{4D} & = & - \left[
\left(\bar{\Gamma}^5_{\,\,\,\beta 5,\gamma} -
\bar{\Gamma}^5_{\,\,\,\beta\gamma, 5}\right) \right. \nonumber \\
&+& \left. \left.\left( \bar{\Gamma}^5_{\,\,\,5 \gamma}
\bar{\Gamma}^5_{\,\,\,\beta 5} + \bar{\Gamma}^5_{\,\,\,\mu\gamma}
\bar{\Gamma}^{\mu}_{\,\,\,\beta 5} +
\bar{\Gamma}^{\alpha}_{\,\,\,5 \gamma }
\bar{\Gamma}^5_{\,\,\,\beta \alpha} - \bar{\Gamma}^5_{\,\,\,55}
\bar{\Gamma}^5_{\,\,\,\beta\gamma} - \bar{\Gamma}^5_{\,\,\,\mu 5}
\bar{\Gamma}^{\mu}_{\,\,\,\beta \gamma} -
\bar{\Gamma}^{\mu}_{\,\,\,5 \mu}
\bar{\Gamma}^5_{\,\,\,\beta\gamma}\right) \right]\right|_{l=l_0},
\label{RW}
\end{eqnarray}
which is symmetric with respect to $\beta \gamma$. The induced
antisymmetric tensor that comes from the contraction of the 5D
Weitzenb\"ock curvature tensor is
$\overbrace{\bar{S}_{\beta\gamma}}^{4D}\equiv
\left.\overbrace{\bar{R}^{a}_{\,\,\,a\beta\gamma
}}^{4D}\right|_{l=l_0}$. This antisymmetric tensor is given by
\begin{equation}\label{mm}
\overbrace{\bar{S}_{\gamma\delta}}^{4D}
=-\left.\left[\bar{\Gamma}^5_{\,\,\,5 \delta,\gamma} -
\bar{\Gamma}^5_{\,\,\,5 \gamma,\delta}\right]\right|_{l=l_0=1/H}.
\end{equation}
To induce the effective 4D scalar Weitzenb\"ock curvature, we must
notice that
\begin{equation}\label{r4}
\overbrace{\bar{R}}^{5D} = g^{ab} \overbrace{\bar{R}_{ab}}^{5D}
\equiv g^{\alpha\beta} \overbrace{\bar{R}_{\alpha\beta}}^{5D} +
g^{55} \overbrace{\bar{R}_{55}}^{5D}=0.
\end{equation}
Hence, if we identify in (\ref{r4}) $\overbrace{\bar{\cal R}}^{4D}
=
\left.g^{\beta\gamma}\,\overbrace{\bar{R}_{\beta\gamma}}^{5D}\right|_{l=l_0}
=h^{\beta\gamma}\,\overbrace{\bar{R}_{\beta\gamma}}^{4D}$ as the
effective 4D induced Weitzenb\"ock scalar curvature we obtain that
this scalar curvature is $\overbrace{\bar{R}}^{4D}=-\left.g^{55}
\overbrace{\bar{R}_{55}}^{5D}\right|_{l=l_0}$ and therefore the
induced 4D Einstein-Weitzenb\"ock equations, finally, are
described by
\begin{eqnarray}
\overbrace{\bar{G}_{\beta\gamma}}^{4D}& = &
\overbrace{\bar{R}_{\beta\gamma}}^{4D} - \frac{1}{2}
h_{\beta\gamma} \overbrace{\bar{\cal R}}^{4D} = -8\pi G \,
\underbrace{\overbrace{\bar{T}_{\beta\gamma}}^{4D}}_{(sym)}, \label{EW}
\\
\overbrace{\bar{S}_{\beta\gamma}}^{4D} &=& - 8\pi G\underbrace{\overbrace{\bar{T}_{\beta\gamma}}^{4D}}_{(ant)}, \label{EW2} \\
\bar{R}_{a 5} & = & 0, \label{ew1} \\
\bar{S}_{a 5} & = & 0, \label{ew2}
\end{eqnarray}
where
$\underbrace{\overbrace{\bar{T}_{\beta\gamma}}^{4D}}_{(sym)}={1\over
2}
\left(\overbrace{\bar{T}_{\beta\gamma}}^{4D}+\overbrace{\bar{T}_{\gamma\beta}}^{4D}\right)$
is the symmetrized energy-momentum tensor in the Weitzenb\"ock
representation and
$\underbrace{\overbrace{\bar{T}_{\beta\gamma}}^{4D}}_{(ant)}={1\over
2}
\left(\overbrace{\bar{T}_{\beta\gamma}}^{4D}-\overbrace{\bar{T}_{\gamma\beta}}^{4D}\right)$
the antisymmetrized contribution. These equations provide the
effective 4D relativistic dynamics. In particular, Eq. (\ref{EW})
describes the dynamics and Eqs. (\ref{ew1})-(\ref{ew2}) provide us
with the constraint conditions.

\subsection{Effective 4D Levi-Civita representation from the induced 4D
Weitzenb\"ock representation}

In order to obtain the effective 4D Einstein equations in the
Levi-Civita representation, we shall use the expression for the
connections (\ref{con}) in Eq. (\ref{ind}). Then we obtain the
effective 4D Ricci tensor
\begin{eqnarray}
\overbrace{{R}_{\beta\gamma}}^{4D} &\equiv
&\left.{\Gamma}^{\alpha}_{\,\,\,\beta\alpha,\gamma} -
{\Gamma}^{\alpha}_{\,\,\,\beta\gamma,\alpha} +
{\Gamma}^\epsilon_{\,\,\,\beta\alpha}
{\Gamma}^{\alpha}_{\,\,\,\epsilon\gamma} -
{\Gamma}^{\epsilon}_{\,\,\,\beta\gamma} {\Gamma}^{\alpha}_{\,\,\,\epsilon\alpha}\right|_{l=l_0} \nonumber \\
&=& \overbrace{\bar{R}_{\beta\gamma}}^{4D} +
\left.\bar{K}^{\alpha}_{\,\,\,\beta\alpha,\gamma} -
\bar{K}^{\alpha}_{\,\,\,\beta\gamma,\alpha} +
\bar{K}^\epsilon_{\,\,\,\beta\alpha}
\bar{K}^{\alpha}_{\,\,\,\epsilon\gamma} -
\bar{K}^{\epsilon}_{\,\,\,\beta\gamma}
\bar{K}^{\alpha}_{\,\,\,\epsilon\alpha}\right|_{l=l_0},
\label{indLC}
\end{eqnarray}
which depends on the effective 4D Ricci tensor in the
Weitzenb\"ock representation given by Eq. (\ref{RW}) and the
contortion $\bar{K}^{\alpha}_{\,\,\,\beta\gamma}$. Finally, the
Einstein equations for a scalar field in the Levi-Civita
representation are
\begin{eqnarray}
\overbrace{{G}_{\beta\gamma}}^{4D}& = &
\overbrace{{R}_{\beta\gamma}}^{4D} - \frac{1}{2} h_{\beta\gamma}
\overbrace{{\cal R}}^{4D} \equiv  -8\pi G \,
\underbrace{\overbrace{{T}_{\beta\gamma}}^{4D}}_{(sym)},
\label{EWW}
\\
\overbrace{{S}_{\beta\gamma}}^{4D}& \equiv  &
\left.\left[\bar{K}^a_{\,\,\,a \gamma,\beta} -
\bar{K}^a_{\,\,\,a \beta,\gamma}\right]\right|_{l=l_0=1/H}=-8 \pi G \,\underbrace{\overbrace{{T}_{\beta\gamma}}^{4D}}_{(ant)}, \label{EW1}
\end{eqnarray}
with $\bar{R}_{a5}=0$ and $\bar{S}_{a5}=0$. On the other hand
$\underbrace{\overbrace{{T}_{\beta\gamma}}^{4D}}_{(ant)}$ is the
effective 4D induced antisymmetric energy-momentum tensor in the
Levi-Civita representation. On the other hand
$\underbrace{\overbrace{{T}_{\beta\gamma}}^{4D}}_{(sym)}$ is the
effective 4D induced symmetric energy-momentum tensor in the
Levi-Civita representation, which invariant of form with respect
to $\underbrace{\overbrace{\bar{T}_{\beta\gamma}}^{4D}}_{(sym)}$
in terms of the scalar fields and their covariant
derivatives\footnote{Notice that for fermionic fields (for
instance with spin $1/2$), the Cartan equation (\ref{EW1}) should
be
\begin{displaymath} \overbrace{{S}_{\beta\gamma}}^{4D} -
\frac{1}{2} \sigma_{\beta\gamma} S =-8 \pi G
\,\underbrace{\overbrace{{T}_{\beta\gamma}}^{4D}}_{(ant)},
\end{displaymath}
where $S= \sigma^{\mu\nu} S_{\mu\nu}$, $\sigma^{\mu\nu}={1\over 2}
\left[\gamma^{\mu}, \gamma^{\nu}\right]$ and $\gamma^{\mu}$ are
the Dirac matrices. However, this issue is beyond the scope of
this work.}. Equations (\ref{EWW},\ref{EW1}) are the same than
obtained by Cartan\cite{CE} and describe the well-known
Cartan-Einstein formalism. A perfect fluid representation of the
stress tensor components should be
\begin{eqnarray}
\underbrace{\overbrace{{T}_{\beta\gamma}}^{4D}}_{(sym)} & = & \frac{1}{2} (P+\rho) \left\{u_{\beta}, u_{\gamma}\right\}-P g_{\beta\gamma}, \\
\underbrace{\overbrace{{T}_{\beta\gamma}}^{4D}}_{(ant)} & = & \frac{1}{2} (P+\rho) \left[u_{\beta}, u_{\gamma}\right],
\end{eqnarray}
where $\left\{u_{\beta}, u_{\gamma}\right\}$ and $\left[u_{\beta},
u_{\gamma}\right]$ denotes, respectively, the anti-commutator and
commutator between the effective tetra-velocities $u_{\beta}$ and
$u_{\gamma}$. Here, we denote the radiation energy density and
pressure by $\rho$ and $P$, respectively. Of course, this
representation excludes the fermion spinor contributions, which
should be included in a high density fermion system\cite{pop1}.

The cases with $\overbrace{{S}_{\beta\gamma}}^{4D}=0$ and
$\overbrace{{R}_{\beta\gamma}}^{4D}\neq 0$ describe a purely
Einstein formalism, but cases with zero curvature
$\overbrace{{R}_{\beta\gamma}}^{4D}=0$ and nonzero torsion
$\overbrace{{S}_{\beta\gamma}}^{4D}\neq 0$, can be related with
the $f(T)$ theories\cite{FT}.

\section{An example: Pre-big bang collapse}

We consider the 5D Riemann-flat spacetime $dS^2= g_{AB} \,dx^A
dx^B$
\begin{equation}\label{1}
dS^2=\left(\frac{l}{l_0}\right)^2 \left[ dt^2 - e^{-2 t/l_0}
\,dr^2\right] -dl^2,
\end{equation}
such that $l$ is the non-compact extra coordinate and the 3D
spatial coordinates (we shall consider units $c=\hbar=1$) are
considered as cartesian $dr^2= dx^2+dy^2+dz^2$. The metric
(\ref{1}) describes 5D extended universe with constant (and
negative) cosmological parameter, which is contracting with the
time $t$. Furthermore, $dr^2 = dx^2 + dy^2 + dz^2$ is the 3D
Euclidean metric. Since the metric (\ref{1}) is Riemann-flat (and
therefore Ricci-flat), hence it is suitable to describe a 5D
vacuum ($G_{AB}=0$) in the framework of Space-Time-Matter (STM)
theory of gravity. When we make a constant foliation $l=l_0$ on
this metric we obtain an effective 4D collapsing universe with an
asymptotic singular size and an energy density which tends to
infinity. In order to avoid this problem, we shall consider a
different metric $dS^2=g_{ab}\,dy^a dy^b$:
\begin{equation}\label{2}
dS^2=\left(\frac{l}{l_0}\right)^2 \left[dt^2 -
\frac{\psi^2(t)}{\psi^2_0} e^{-2 t/l_0} \,dr^2\right] -dl^2,
\end{equation}
where $\psi(t)= \psi_0\,\cosh{(t/\psi_0)}$ is a function of the
time $t$. The metrics (\ref{1}) and (\ref{2}) are related by the
transformations
\begin{equation}
g_{ab}= e^A_{\,\,b}\, e^B_{\,\,b}\, g_{AB},
\end{equation}
such that $e^A_b$ and its inverse $\bar{e}^a_B$, are given by
\begin{eqnarray}
e^A_b &=& {\bf diag} \left[\left(1,\frac{\psi(t)}{\psi_0},
\frac{\psi(t)}{\psi_0}, \frac{\psi(t)}{\psi_0}, 1\right)\right],
\\
\bar{e}^a_B &=& {\bf diag}
\left[\left(1,\frac{\psi_0}{\psi(t)},\frac{\psi_0}{\psi(t)},\frac{\psi_0}{\psi(t)},
1\right)\right].
\end{eqnarray}
Notice that the metric (\ref{1}) is also Weitzenb\"ock-flat. The
Weitzenb\"ock torsion is zero, so that the Levi-Civita connection
$\Gamma^A_{\,\,\,BC}$ is coincident with the Weitzenb\"ock
connection $^{(W)} \Gamma^A_{\,\,\,BC}$: $\Gamma^A_{\,\,\,BC} =
\bar{\Gamma}^A_{\,\,\,BC}$. On the other hand, the metric
(\ref{2}) is not Riemman-flat $R^a_{\,\,\,bcd} \neq 0$, so that it
is not a good candidate to realize standard IMT. However, from the
point of view of the Weitzenb\"ock geometry, it is
Weitzenb\"ock-flat and also has nonzero torsion: $\bar{
R}^a_{\,\,\,bcd}=0$, $\bar{T}^a_{\,\,\,bc}\neq 0$. In the
spacetime (\ref{2}) the connections are related by the contortion
$K^a_{\,\,\,bc} = \Gamma^a_{\,\,\,bc} -
\bar{\Gamma}^a_{\,\,\,bc}$, with $ \bar{\Gamma}^a_{\,\,\,bc}=
\bar{e}^a_{\,\,A} \partial_c\, e^A_{\,\,\,b}$.

In order to describe the collapse, we shall consider the action on
a Riemann-flat 5D metric (\ref{1}), for a Levi-Civita scalar field
$\varphi(x^A)$, which is free of interactions
\begin{equation}
{\cal I} = \int d^4x \,dl \sqrt{|g_1|} \left(\frac{{\cal
{R}}}{16\pi G} + \frac{1}{2} g^{AB} {\varphi}_{,A} {\varphi}_{,B}
\right),
\end{equation}
where $g_1$ is the determinant of the tensor metric $g_{AB}$. The
equation of motion for $\varphi(x^A)$ on the Riemann-flat and for
$\bar{\varphi}(x^A)$ on the Weitzenb\"ock-flat spacetime
(\ref{1}), are $\nabla^A\,
\varphi_{,A}=\bar{\nabla}^{A}\bar{\varphi}_{,A}=0$, such that
\begin{eqnarray}
\nabla^A \varphi_{,A} &\equiv  & g^{AB}
\left(\partial_A\varphi_{,B} -
\Gamma^C_{\,\,\,AB} \varphi_{,C}\right), \\
\bar{\nabla}^A \bar{\varphi}_{,A} & \equiv  & g^{AB}
\left(\partial_A\varphi_{,B} - \bar{\Gamma}^C_{\,\,\,AB}
\varphi_{,C}\right).
\end{eqnarray}
Since $g^{AB}\,\bar\varphi_{,A} \bar\varphi_{,B} = g^{ab}
\bar\varphi_{,a} \bar\varphi_{,b}$, we can consider that the
action ${\cal I}$ on the metric (\ref{2}) is an invariant after
making the transformation ${x^A} \rightarrow {y^a}$, so that one
can describe the Weitzenb\"ock field $\bar{\varphi}(y^a)$ on
(\ref{2}), as
\begin{equation}
{\cal I} = \int d^4y \,dl \sqrt{|g_2|} \left(\frac{{\cal
\bar{R}}}{16\pi G} + \frac{1}{2} g^{ab} \bar{\varphi}_{,a}
\bar{\varphi}_{,b} \right),
\end{equation}
where $g_2$ is the determinant of the metric tensor $g_{ab}$ and
$\bar{\cal R}=0$ is a Weitzenb\"ock invariant: $g^{ab}
\bar{R}^c_{\,\,\,abc}$. The equation of motion for
$\bar{\varphi}(y^a)$ is given by the null D'Alembertian of
Weitzenb\"ock: $\bar{\nabla}^a \bar{\varphi}_{,a}=0$, or
\begin{equation}
\ddot{\bar{\varphi}} - \frac{ e^{2 t/l_0}}{\cosh^2{(t/l_0)}}
\nabla^2_r \bar{\varphi} - \left(\frac{l}{l_0}\right)^2
\frac{\partial^2 \bar{\varphi}}{\partial l^2} =0.
\end{equation}
This means that the scalar field $\bar{\varphi}(y^a)$ is a free
scalar field in a Weitzenb\"ock representation. In other words,
its origin is not geometric, but their physical properties
(describing a 5D physical vacuum in absence of interactions in a
Weitzenb\"ock representation) have a geometric dependence. On the
other hand, from the point of view of the Levi-Civita
representation, this means that
\begin{equation}
{\nabla}^a {\varphi}_{,a} =\bar{\nabla}^a {\varphi}_{,a}- g^{ab}
K^c_{\,\,\,ab} \varphi_{,c}=0,
\end{equation}
where we have made use of the fact that $\bar{\nabla}^a
{\bar\varphi}_{,a}={\nabla}^a {\varphi}_{,a}=0$. The relevant
contortion components are
\begin{eqnarray}
K^1_{\,\,22} & = &\frac{e^{-2t/l_0}}{l_0} \cosh{(t/l_0)} \left[
\sinh{(t/l_0)} -\cosh{(t/l_0)}\right], \qquad K^1_{\,\,55}=0, \\
K^5_{\,\,11}& = & l/l^2_0, \qquad K^5_{\,\,22}=- \frac{l}{l^2_0}
e^{-2t/l_0} \cosh^2{(t/l_0)}.
\end{eqnarray}
Hence, the equation of motion for the Levi-Civita scalar field
$\varphi(y^a)$, finally, as a result is found to be
\begin{equation}
\ddot\varphi + \frac{3}{l_0} \left[\tanh{(t/l_0)} -1\right]
\dot\varphi - \frac{ e^{2 t/l_0}}{\cosh^2{(t/l_0)}} \nabla^2_r
{\varphi} - \left(\frac{l}{l_0}\right)^2 \left[ \frac{\partial^2
{\varphi}}{\partial l^2}+\frac{4}{l}
\frac{\partial\varphi}{\partial l}\right] =0.
\end{equation}
This field can be written as a Fourier expansion in terms of the
modes $\varphi_{k}(y^a) = {\bf A_k} \,\xi_k(t) \, e^{i
\vec{k}.\vec{y}}\, \Lambda(l)$, where $\Lambda(l)$ and $\xi_k(t)$
are given, respectively, by the solutions of the equations
\begin{eqnarray}
&& \left(\frac{l}{l_0}\right)^2 \left[ \frac{d^2
{\Lambda}}{d l^2}+\frac{4}{l}
\frac{d\Lambda}{d l}\right]= M^2\, \Lambda(l), \label{e1} \\
&& \ddot\xi_k + \frac{3}{l_0} \left[\tanh{(t/l_0)} -1\right]
\dot\xi_k +\left[ \frac{ e^{2 t/l_0}}{\cosh^2{(t/l_0)}} k^2 -
M^2\right] \xi_k(t) =0, \label{e2}
\end{eqnarray}
where $M^2$ is a separation of variables constant and de dot
denotes the derivative with respect to $t$. The solutions for Eqs.
(\ref{e1}) and (\ref{e2}) are
\begin{eqnarray}
\Lambda(l) &= &\left(\frac{l}{l_0}\right)^{-3/2} \left[ \Lambda_1
\left(\frac{l}{l_0}\right)^{\frac{\sqrt{9+4M^2l^2_0}}{2}} +
\Lambda_2
\left(\frac{l}{l_0}\right)^{-\frac{\sqrt{9+4M^2l^2_0}}{2}}
\right], \label{lam}\\
\xi_k(t)&=& \left[ 1+ \tanh{(t/l_0)}\right] \left[\tanh{(t/l_0)}
-1\right]^{\frac{\sqrt{M^2-4 k^2}}{2}}
\nonumber \\
&\times & \left\{ A_1 \,{\bf _2F_1}\left[ [a_1,b_1], c_1,
2^{\frac{-\left(2+\sqrt{9+M^2l_0^2}\right)}{2}} \left[ 1+
\tanh{(t/l_0)}\right]^{\frac{3-\sqrt{9+M^2l^2_0}}{2}}\right]
\right.\nonumber \\
&+& \left. B_1 \, {\bf _2F_1}\left[ [a_2,b_2], c_2,
2^{\frac{\left(\sqrt{9+M^2l_0^2}-1\right)}{2}} \left[ 1+
\tanh{(t/l_0)}\right]^{\frac{3+\sqrt{9+M^2l^2_0}}{2}}\right]\right\},
\end{eqnarray}
where $(A_1,A_2)$ are constants and ${\bf _2F_1}$ denotes the
Gaussian hypergeometric function with parameters
\begin{eqnarray}
a_{1,2} & = & \frac{l_0}{2} \sqrt{M^2- 4k^2} + \sqrt{1-l^2_0 k^2}
\mp \frac{\sqrt{l^2_0 M^2+9} +1}{2}, \\
b_{1,2} & = & \frac{l_0}{2} \sqrt{M^2- 4k^2} - \sqrt{1-l^2_0 k^2}
\mp \frac{\sqrt{l^2_0 M^2+9} +1}{2}, \\
c_{1,2} &=& \left[1 \mp \sqrt{9+ l^2_0 M^2}\right].
\end{eqnarray}
In the ultraviolet (UV) limit of the spectrum, for which $k\gg M$
and $k l_0 \gg 1$, these parameters take the asymptotic form
\begin{eqnarray}
a_{1,2} & \simeq & 2 i \,l_0 k \mp \frac{\sqrt{l^2_0 M^2+9} +1}{2}, \\
b_{1,2} & \simeq & \mp \frac{\sqrt{l^2_0 M^2+9} +1}{2}, \\
c_{1,2} &=& \left[1 \mp \sqrt{9+ l^2_0 M^2}\right].
\end{eqnarray}
When $t\rightarrow \infty$, one can see that
$\tanh{(t/l_0)}\rightarrow 1$, so that the equation of motion
(\ref{e2}) tends asymptotically to
\begin{equation}
\ddot\xi_k  +\left[4 k^2 -
M^2\right] \xi_k(t) =0,
\end{equation}
which has the asymptotic solution
\begin{equation}\label{flu}
\left.\xi_k(t)\right|_{t\rightarrow \infty} \rightarrow A(k)\, e^{i
\,\sqrt{4k^2-M^2 } t} + B(k) \,e^{-i \,\sqrt{4k^2-M^2 } t}.
\end{equation}
This means that the amplitude of the fluctuations remains
asymptotically constant as $t \rightarrow \infty$.

\subsection{Effective 4D dynamics}

Now we consider a constant foliation $l=l_0=1/H$ on the fifth
coordinate. The effective 4D metric for observers that move with
Weitzenb\"ock velocities $\bar{u}^a=(1,0,0,0,0)$, will be $dS^2 =
h_{\alpha\beta} dy^{\alpha} dy^{\beta}$, or
\begin{equation}\label{mett}
dS^2=\left.\left(\frac{l}{l_0}\right)^2 \left[dt^2 -
\frac{\psi^2(t)}{\psi^2_0} e^{-2 t/l_0} \,dr^2\right]
-dl^2\right|_{l=1/H} \rightarrow dS^2 = dt^2 - \cosh^2{(H_0 t)}
e^{-2H_0t} dr^2.
\end{equation}
Here, the scale factor of the universe which collapses is $a(t) =
\cosh{\left(t/l_0\right)} \, e^{-l^{-1}_0 t}$, the Hubble
parameter $H(t)={\dot{a}\over a}$ and the deceleration parameter
$q=-{\ddot{a} a\over \dot{a}^2}$. This is a very interesting
behavior because the model describes a contracting universe which
has an asymptotic finite size $a_{min}=1/(2H_0)$. This effect is
due to the fact that in General Relativity the action is invariant
under time reflections. Thus, to any standard cosmological
solution $H(t)$, describing decelerated expansion and decreasing
curvature ($H > 0$, $\dot{H} < 0$), time reversal associates a
"reflected" solution, $H(-t)$, describing a contracting Universe.
In a string cosmology context, these solutions are called
dual\cite{gv}. In this work we are dealing with an extra
dimensional cosmological model where the extra dimension is
non-compact. However, this duality is preserved and the
interpretation of the results obtained by Gasperini and Veneziano
in \cite{gv} are preserved. A possible cosmological application of
such a cosmological scenario is bouncing cosmology, in which the
cosmic singularity ($a=0$) is avoided due to the repulsive effects
produced by fermions during the collapse, which are more
significant at very short distances\cite{pop}.

In our example $\overbrace{\bar{S}_{\beta\gamma}}^{4D}=0$, so that
the relativistic dynamics on the effective 4D metric (\ref{mett}),
being given by Eqs. (\ref{EWW}) and (\ref{EW1}). The physical
information is provided by the effective 4D energy-momentum tensor
in the Levi-Civita representation:
\begin{equation}
\overbrace{{T}_{\beta\gamma}}^{4D} = \left.{\varphi}_{,\beta}
{\varphi}_{,\gamma} - g_{\beta\gamma} \left[ \frac{1}{2}
g^{\sigma\mu} {\varphi}_{,\sigma} {\varphi}_{,\mu}+\frac{1}{2}
g^{55} \left({\varphi}_{,5}\right)^2 \right]\right|_{l=l_0=1/H_0}.
\end{equation}
Notice that
$\left.g^{\alpha\beta}\right|_{l=l_0=1/H_0}=h^{\alpha\beta}$.
Here, the last term in the brackets can be identified with the
effective 4D scalar potential in the Levi-Civita representation:
$V({\varphi})\equiv -\left.\frac{1}{2} g^{55}
\left({\varphi}_{,5}\right)^2\right|_{l=l_0=1/H_0}$. Using Eqs.
(\ref{EW}), (\ref{EW1}), (\ref{ew1}), (\ref{ew2}), we obtain the
effective 4D relativistic dynamics. In our example
$\overbrace{S_{\beta\gamma}}^{4D} =0$, which agrees with what one
expects for a scalar field because the antisymmetric contribution
of the effective 4D stress tensor becomes null:
$\underbrace{\overbrace{T_{\beta\gamma}}^{4D}}_{(ant)} =0$. The
effective 4D symmetric contribution of the energy-momentum tensor
components are
\begin{eqnarray}
\overbrace{T^0_{\,\,\,0}}^{4D} &=& \frac{\dot\varphi^2}{2} +
\frac{1}{2
a^2} \left(\vec{\nabla}\varphi\right)^2 + V(\varphi), \\
\overbrace{T^i_{\,\,\,j}}^{4D} &=&- \left[\frac{\dot\varphi^2}{2}
- \frac{1}{6 a^2} \left(\vec{\nabla}\varphi\right)^2 -
V(\varphi)\right] \delta^i_{\,\,\, j},
\end{eqnarray}
where
\begin{equation}\label{pot}
V(\varphi) = \left.{1\over 2} \left[ \frac{d}{dl} {\bf
ln}\left[\Lambda\right] \right]^2 \varphi^2\right|_{l_0}.
\end{equation}
In other words, the effective 4D squared mass of the scalar field
is related to the extra dimensional static solution of $\varphi$
by the expression\footnote{In the cases where $M_{eff}^2 >0$ and
the universe expands with a nearly constant energy density given
by $\rho \simeq \left< V(\varphi)\right>$ and a pressure $P \simeq
-\left<V(\varphi)\right>$, the potential (\ref{pot}) is a good
candidate to describe inflation.}
\begin{equation}\label{mass}
M_{eff}^2 = \left.\left[ \frac{d}{dl} {\bf ln}\left[\Lambda\right]
\right]^2\right|_{l=l_0}.
\end{equation}
Finally, the effective 4D Einstein equations in the Levi-Civita
representation, are (we use 3D cartesian coordinates and the
foliation $l=l_0=1/H_0$)
\begin{eqnarray}
{G}^0_{\,\,0} & = & -\frac{3 H_0^2}{\cosh^2{(H_0 t)}}
\left[\cosh{(H_0 t)} -\sinh{(H_0 t)}\right]^2, \label{30} \\
{G}^i_{\,\,j} & = & -\frac{H_0^2}{\cosh^2{(H_0 t)}}
\left[\cosh{(H_0 t)}-\sinh{(H_0 t)}\right] \left[ 5\cosh{(H_0 t)}
- \sinh{(H_0 t)}\right] \delta^i_{\,\,j}, \label{31}
\end{eqnarray}
so that, using the fact that the Einstein equations are,
respectively, $G^{0}_{\,\,0} = -8\pi G\,\rho $ and $G^x_{\,\,x} =
G^y_{\,\,y}= G^z_{\,\,z}= 8\pi G \, P$, we obtain the equation of
state for the universe
\begin{equation}\label{state}
\frac{P}{\rho} = \omega(t) = - \frac{1}{3} \frac{\left[ 5
\cosh{(H_0 t)} - \sinh{(H_0 t)} \right]}{\left[\cosh{(H_0 t)}
-\sinh{(H_0 t)}\right]}.
\end{equation}
Notice that $\omega$ always remains with negative values
$\omega(t) <-1$, and evolves from $-5/3$ to $-\infty$, for large
asymptotic times. The effective 4D scalar curvature
\begin{equation}
{{{R}}} = \frac{6 H_0^2}{\cosh^2{(H_0 t)}} \left[\cosh{(H_0
t)}-\sinh{(H_0 t)}\right] \left[ 3\cosh{(H_0 t)} - \sinh{(H_0
t)}\right],
\end{equation}
decreases with the time and has a null asymptotic value
$\left.{{R}}\right|_{t\rightarrow \infty} \rightarrow 0 $. We are
considering a spatially isotropic and homogeneous background, so
that we shall consider an averaging value with respect to a
Gaussian distribution on a Euclidean 3D volume. The late time
expectation values for both, the energy density and the pressure,
are zero [see Eqs. (\ref{30}) and (\ref{31})]
\begin{eqnarray}
\left.{\rho}\right|_{t\rightarrow \infty} = \left<0|{T}^0_{\,\,0}
|0\right> & = & \left< \left[\frac{1}{2}\dot{{\varphi}}^2 +
\frac{1}{2 a^2(t)} \left(\vec{\nabla} {\varphi}\right)^2\right]+
\frac{1}{2}
 M_{eff}^2 \varphi^2 \right>_{l=1/H_0}=0, \label{34} \\
\left.{P}\right|_{t\rightarrow \infty}  = -\left<0|
{T}^i_{\,\,j}|0\right> & = &  \delta^i_{\,\,j} \left< \left[
\frac{1}{2} \dot{{\varphi}}^2 - \frac{1}{6 a^2(t)}
\left(\vec{\nabla} {\varphi}\right)^2\right] - \frac{1}{2}
 M_{eff}^2 \varphi^2  \right>_{l=1/H_0}=0,
\label{35}
\end{eqnarray}
 where we denote by $\left<0| ... |0\right>$ as the quantum expectation value
calculated on a 4D vacuum state.

\subsection{Particular solution: asymptotic collapsing state}

In order to obtain the effective 4D squared mass $M^2 _{eff}$ we
must calculate the asymptotic energy density, which is zero
because $G^0_{\,\,0} \rightarrow 0$ for late times (see Eq.
(\ref{30}). The effective 4D scalar field
$\varphi(y^{\alpha},l=l_0)$ can be written as a Fourier expansion
on the 4D hypersurface
\begin{equation}
\varphi(y^{\alpha},l_0) = \frac{\Lambda(l_0)}{(2\pi)^{3/2}} \int d^3 k \left[ {\bf A}_k \, e^{i \vec k.\vec y} \,\xi_k(t) + h.c.\right],
\end{equation}
such that the operators of creation and destruction comply with
the algebra
\begin{equation}
\left[{\bf A}_k, {\bf A}_{k'}^{\dagger}\right] = \delta^{(3)}(\vec k - \vec k'), \qquad
\left[{\bf A}_k^{\dagger}, {\bf A}_{k'}^{\dagger}\right]=\left[{\bf A}_k, {\bf A}_{k'}\right]=0.
\end{equation}
Hence the canonical structure $\left[ \varphi(t,\vec y,l_0),
\dot\varphi(t,\vec{ y'},l_0)\right]= i\, a^{-3}\, \delta^{(3)}
\left(\vec y - \vec{y'}\right)$, will be ensured when the
condition of normalization
\begin{equation}
\xi_k \dot\xi_k^* - \dot\xi_k \xi_k^* = \frac{i}{a^3(t)},
\end{equation}
is fulfilled. To make this happen on the asymptotic late-time
state, we must require that
\begin{equation}\label{nor}
\left.\xi_k \dot\xi_k^* - \dot\xi_k \xi_k^*\right|_{t\rightarrow \infty} = \frac{i}{a^3_{min}},
\end{equation}
in Eq. (\ref{flu}). A particular solution of the condition
(\ref{nor}) is given by $A(k)=0$ and $B(k)={2 H^{3/2}_0 \over
\left[4 k^2 - M^2\right]^{1/4}}$. With this solution we find that
the asymptotic energy density is
\begin{equation}
\left.{\rho}\right|_{t\rightarrow \infty} = \frac{H^3_0}{\pi^2} \int^{4\pi H_0}_0 dk\,\frac{k^2}{\left[4 k^2 - M^2\right]^{1/2}}\,
\left[\frac{(4k^2- M^2)}{2} + 2 k^2 + \frac{M^2_{eff}}{2}\right]
=0, \label{in}
\end{equation}
where the limits of integration are
$(k_{min}=0,\,k_{max}=2\pi/a_{min}=4\pi H_0)$, such that the
minimum scale factor is given by the asymptotic scale factor
during the collapse: $a_{min}= 1/(2H_0)$. On the other hand, from
Eq. (\ref{31}) we know that the asymptotic pressure is zero, so
that from the expression (\ref{35}), we obtain
\begin{equation}
\left.{P}\right|_{t\rightarrow \infty} = \frac{H^3_0}{\pi^2} \int^{4\pi H_0}_0 dk\,
\frac{k^2}{\left[4 k^2 - M^2\right]^{1/2}}\,
\left[\frac{(4k^2- M^2)}{2} - \frac{2}{3} k^2-
\frac{M^2_{eff}}{2}\right] =0. \label{in1}
\end{equation}
From Eqs. (\ref{in}) and (\ref{in1}), we obtain the effective 4D
mass of the potential $V(\varphi)$ and the parameter $M^2$
\begin{equation}\label{mm}
M^2_{eff} =-32 \pi^2 H^2_0,\qquad M^2= 64\pi^2 H^2_0.
\end{equation}
This means that the effective 4D potential
$\left<0|V(\varphi)|0\right> <0$ is negative and the asymptotic
system is unstable. The system which we are describing is very
similar to a gravitational collapse suffered by an observer in a
Schwarzschild black-hole, but in our case there is no cosmic
singularity, because $a_{min} \neq 0$. Notice that the asymptotic
values of radiation energy density and pressure are zero, because
all the matter is eaten by the black-hole. To make a more
realistic model one would include the contribution of fermions as
a condensate, in order to describe the physical origin of
repulsion on very small scales. Of course the study of such a
physical problem deserves a more intense study, but it goes beyond
the scope of this paper.

\section{Final remarks}

We have extended the theoretical background for the IMT of
gravity. The fact that one can define a 5D vacuum on a Levi-Civita
5D curved spacetime using the Weitzenb\"ock representation opens a
huge window of possible applications for this theory. In this
framework we study a pre-big bang collapsing universe that ends
with finite size and energy density. Notice that we have
restricted our analysis to a 5D coordinate basis, but the
formalism can be extended to a non-coordinate basis. As an example
we have studied a pre-big bang collapsing universe. The difference
with respect to earlier studies\cite{mb} lies in the fact that the
scalar field induced in the Levi-Civita representation is massive,
its squared mass being given by the expression (\ref{mass}). We
have found that the effective 4D expectation value for the
potential is quadratic and negative, so that the asymptotic late
time collapse is an unstable system with an equation of state
$\omega = -\infty$. This suggests a good initial state for a new
big-bang.

\section*{Acknowledgements}

\noindent The authors acknowledge P. S. Wesson for his interesting
comments. Furthermore, we acknowledge UNMdP and CONICET Argentina
for financial support.

\end{document}